\documentclass[12pt]{article}
\usepackage{graphicx}
\newfont{\larom}{cmbx10 scaled\magstep3}
\newfont{\bsan}{cmssbx10}
\begin{document}

\begin{center}
  {\larom On Modelling a Relativistic Hierarchical (Fractal) Cosmology by 
   Tolman's Spacetime . II. Analysis of the Einstein--de Sitter Model}  

  \vspace{10mm}
  {\Large Marcelo B. Ribeiro \\}
  \vspace{5mm}
  {\normalsize Astronomy Unit \\ School of Mathematical
  Sciences \\ Queen Mary and Westfield College \\ Mile End Road \\
  London E1 4NS \\ England}



  \vspace{10mm}
  {\bf ABSTRACT}
\end{center}
  \begin{quotation}
    \small

    This paper studies the spatially homogeneous Einstein--de Sitter
    cosmological model in the
    context of a relativistic hierarchical (fractal) cosmology as
    developed in paper I. We treat the Einstein--de Sitter model as
    a special case of Tolman's spacetime, obtained by the appropriate
    choice of the latter's three arbitrary functions. We calculate the
    observational relations along the past light cone of the model under
    consideration and carry out an investigation of whether or not it
    has fractal behaviour. We have found that the Einstein--de Sitter model
    does not seem to remain homogeneous along the geodesic and that it
    also has no fractal features along the backward null cone.  

    \vspace{3mm}

    \underline{Subject headings}: cosmology: theory

  \end{quotation}

\vspace{6mm}
\newpage

\section{Introduction}
  ~~~~~The recent  galactic redshift surveys showing a very
  inhomogeneous picture for the distribution of galaxies (de Lapparent,
  Geller \& Huchra 1986; Saunders et al. 1991) have stimulated the
  trend of study of the galactic clustering problem which assumes that
  the large scale structure of the universe can be described as being a
  self--similar fractal system. Under this philosophy there have been
  studies in which the distribution of galaxies is assumed to form a
  multifractal system (Jones et al. 1988; Mart\'{\i}nez et al. 1990),
  but the first single fractal Newtonian model advanced as a description
  of the large scale structure is due to Pietronero (1987).

  ~~~~~This paper is the second of a series where we develop a
  relativistic fractal cosmology obtained by the relativistic
  generalization of Pietronero's (1987) model. In paper I (Ribeiro 1992)
  we have argued that the all sky redshift surveys mentioned above
  present observations consistent both with the old Charlier's
  hypothesis of hierarchical clustering and with fractals, where the
  latter is basically a more precise conceptualization of the
  scaling idea implicit in the
  hierarchical clustering hypothesis. Also in paper I we have used
  Pietronero's basic hypothesis to propose similar ones in a relativistic
  framework and to develop observational relations compatible with
  fractals in Tolman's spacetime.

  ~~~~~The purpose of this paper is to apply those observational
  relations to the specific and analytically manageable special case of
  the Einstein--de Sitter model, which is obtained by a particular choice of
  the three arbitrary functions of Tolman's spacetime. Our basic goal
  here is to answer the
  question of whether or not the Einstein--de Sitter model is compatible
  with a fractal description of galactic clustering as developed in
  paper I, and we intend to do so by studying the consequences of paper
  I's observational relations in a Einstein--de Sitter universe. We
  believe that this problem must be addressed because Einstein--de
  Sitter is spatially homogeneous and it is not at all clear what would
  be the behaviour of its fractal observational relations as we go down the
  geodesic, through hypersurfaces of
  constant $t$ each having a different value for the proper density.
  The calculations shown next also serve as an illustration of the
  theory developed in paper I.

  ~~~~~The plan of the paper is as follows. In \S 2 we describe
  very briefly the Tolman observational relations presented in paper I and
  discuss how the Tolman metric relates to its Friedmann counterpart. In
  \S 3 we apply these observational relations to the Einstein--de Sitter case
  and analyse the results. There is a conclusion in \S 4.

\section{Observational Relations in Tolman's Spacetime}
  ~~~~~The Tolman (1934) metric for the motion of spherically symmetric
  dust is
  \begin{equation}
   dS^2=dt^2-\frac{{R'}^2}{f^2} dr^2 - R^2 (d \theta^2+\sin^2 \theta d \phi^2).
   \label{p1}
  \end{equation}
  The Einstein field equations (with $c=G=1$ and $\Lambda=0$) for metric
  (\ref{p1}) reduce to a single equation
  \begin{equation}
   2 R {\dot{R}}^2 + 2 R (1-f^2) = F,
   \label{p2}
  \end{equation}
  where a dot means $\partial/\partial t$ and a prime means
  $\partial/\partial r$; $f(r)$ and $F(r)$ are two arbitrary functions
  and $R(r,t) \geq 0$. For $f^2 = 1$ the field equation (\ref{p2}) has
  the following solution:
  \begin{equation}
   R=\frac{1}{2}{(9F)}^{1/3}{(t+\beta)}^{2/3}
   \label{p3}
  \end{equation}
  where $\beta(r)$ is a third arbitrary function which gives the local
  time passed since the singularity surface, that is, since the big
  bang. Equation (\ref{p2}) also has solutions for $f^2 < 1$ and $f^2 >
  1$, but they are of no interest for us here (see paper I and
  references therein). The local density is given by
  \begin{equation}
  8 \pi \rho = \frac{F'}{2 R' R^2}
  \label{p4}
  \end{equation}
  and if we adopt the radius coordinate $r$ as the parameter along the
  backward null cone, we can then write the radial null geodesic of
  metric (\ref{p1}) as
  \begin{equation}
  \frac{dt}{dr} = - \frac{R'}{f}.
  \label{p5}
  \end{equation} 

  ~~~~~For the sake of clarity of this work, we shall briefly describe
  the relationship of the equations above to the usual Friedmann
  spacetime. The junction conditions between the Tolman and Friedmann
  metrics calculated in paper I show that in order to obtain the latter
  from the former we have to assume that
   \begin{equation}
    R(r,t) = a(t) \ g(r), \ \ \ \ f(r) = g'(r).
    \label{A1}
  \end{equation}
  By substituting equations (\ref{A1}) into the metric (\ref{p1}) we get
  \begin{equation}
   dS^2=dt^2-a^2(t) \left\{ dr^2 - g^2(r) \left[ d \theta^2+\sin^2 \theta
	d \phi^2 \right] \right\},
   \label{A2}
  \end{equation}
  which is a Friedmann metric if
  \begin{equation}
    g(r) = \left\{ \begin{array}{ll}
	           \sin r,     &   \\
		      r,       &   \\
	           \sinh r.    &   
	           \end{array}
           \right.
     \label{A3}
  \end{equation}
  If we now substitute equations (\ref{A1}) in equation (\ref{p4}) and
  integrate it, that gives
  \begin{equation}
   \frac{F}{4} = \frac{4 \pi}{3} \rho a^3 g^3.
   \label{A4}
   \end{equation}
   It is worth noting that the time derivative of the equation
   above gives the well-known relation for the matter-dominated era of
   a Friedmann universe:
   \[
     \frac{d}{dt} \left( \rho a^3 \right) = 0.
   \]
   Equation (\ref{A4}) is necessary in order to show the physical role
   played by $F(r)$ and to deduce the usual Friedmann equation.
   This is possible by substituting equations (\ref{A1}) and (\ref{A4})
   into equation (\ref{p2}). The result may be written as
   \begin{equation}
    \dot{a}^2 = \frac{8 \pi}{3} \rho a^2 - K 
   \label{A5}
   \end{equation}
   where
   \[
     K = \frac{1-{g'}^2}{g^2}.
   \]
   It is easy to see that $K=+1, \ 0, \ -1$ if $g=\sin r, \ r, \ \sinh
   r$, respectively, and this shows that equation (\ref{A5}) is indeed the
   usual Friedmann equation. Let us now write equation (\ref{A5}) in the
   form
   \begin{equation} 
       \frac{\dot{a}^2 g^2}{2} - \frac{m}{ag} = - \left(1-{g'}^2 \right)
    \label{A7}
   \end{equation} 
   where
   \begin{equation}
    m(r) = \frac{4 \pi}{3} \rho a^3 g^3.
    \label{A8}
   \end{equation}
   Equation (\ref{A7}) is interpreted as an energy equation and, in
   consequence, $m(r)$ is the gravitational mass inside the coordinate
   $r$. Thus $4 m(r) = F(r)$ and this clarifies the role of the
   function $F(r)$ in providing the gravitational mass of the system. In
   addition, equation (\ref{A7}) shows that the function $f(r)$ in Tolman's
   spacetime gives the total energy of the system (see also paper I, \S 4).

   ~~~~~Finally, a word should be said about the function $\beta(r)$
   that gives the big bang time. If ``now'' is defined as $t=0$ and if
   $\beta(r) = 0$, then the hypersurface $t=0$ is singular, that is,
   $R=0$ everywhere \footnote{This is also valid for $f^2 > 1$ and $f^2
   < 1$ type solutions of equation (\ref{p2}). See paper I, \S 3, for
   the analytical expressions of $R$ in these two cases.}. So $\beta(r)$ 
   gives the age of the universe which
   in Tolman's spacetime may change if different observers are situated
   at different radial coordinates $r$. This is a remarkable departure
   from Friedmann's model that gives the same age of the universe for
   all observers on a hypersurface of constant $t$. In other words, in
   a Friedmann universe the big bang is simultaneous while in a Tolman one
   it may be non-simultaneous, that is, the big bang may have occurred
   at different proper times in different locations. As a consequence 
   another essential ingredient in reducing the Tolman metric to
   Friedmann is $\beta =$ constant, and so the linkage
   between $\beta$ and the Hubble constant is of the form
   \begin{equation}
    \frac{\dot{R}}{R} = \frac{\dot{a}}{a} = H(t), \ \ \ \mbox{for} \ \ 
    \beta = \beta_0,
    \label{A9}
   \end{equation}
   where $\beta_0$ is a constant. Considering equation (\ref{p3}) it is
   straightforward to conclude that
   \begin{equation}
    \beta_0 = \frac{2}{3 H_0},
   \label{A10}
   \end{equation}
   where $H_0 = H(0)$. Equation (\ref{A10}) gives the relationship 
   between $\beta_0$ and the Hubble constant $H_0$ in a Einstein-de
   Sitter universe.

  ~~~~~Let us now return to the discussion of the metric (\ref{p1}) and
  its observational relations. It was shown in paper I that in Tolman's
  spacetime the redshift may be written as
  \begin{equation}
  1+z = {(1-I)}^{-1}
  \label{p6}
  \end{equation}
  where the function $I(r)$ is the solution of the differential equation
  \begin{equation}
  \frac{dI}{dr} = \frac{\dot{R'}}{f} (1-I).
  \label{p7}
  \end{equation}
  The luminosity distance $d_l$ and the cumulative number count $N_c$
  are given by
  \begin{equation}
   d_l = R{(1+z)}^2,
   \label{p8}
  \end{equation}
  \begin{equation}
  N_c=\frac{1}{4 M_G} \int \frac{F'}{f} dr,
  \label{p9}
  \end{equation}
  where $M_G$ is the average galactic rest mass ($ \sim 10^{11}
  M_{\odot}$). The volume $V$ of the sphere which contains the sources,
  and the volume density (average density) $\rho_v$, have the form
  \begin{equation}
   V=\frac{4}{3} \pi {(d_l)}^3,
   \label{p10}
  \end{equation}
  \begin{equation}
   \rho_v = \frac{N_c M_G}{V}.
   \label{p11}
  \end{equation} 
  The proposed relativistic version of Pietronero's (1987) generalized
  mass-length relation is given by 
  \begin{equation}
   N_c = \sigma {(d_l)}^D
   \label{p12}
  \end{equation}    
  where $\sigma$ is a constant related to the lower cutoff of the
  fractal system and $D$ is its fractal dimension. If we substitute
  equations (\ref{p10}) and (\ref{p12}) into equation (\ref{p11}) we
  get de Vaucouleurs' density power law
  \begin{equation} 
   \rho_v = \frac{3 \sigma M_G}{4 \pi} {(d_l)}^{- \gamma}, \ \ \ \
   \gamma = 3-D.
   \label{p13}
  \end{equation}  

\section{Application to the Einstein--de Sitter Case}
  ~~~~~As discussed in the previous section, the Einstein--de Sitter
  model can be obtained from Tolman's spacetime as a special case
  when the three arbitrary functions take the form
\begin{equation}
 f(r) = 1, \ \ \ F(r) = \frac{8}{9} r^3, \ \ \ \beta(r) = \beta_0,
 \label{p14}
\end{equation}
where $\beta_0$ is a constant. The solution (\ref{p3}) of the field
equation and its derivatives for this case may be seen below:
\begin{equation}
 \left\{ \begin{array}{ll}
	 R(r,t) = r {(t+\beta_0)}^{2/3}; &  \dot{R}(r,t) =
	 { \mbox{$ 2 $}\over \mbox{$ 3 $} } r {(t+\beta_0)}^{-1/3}; \\
	 ~~~  &  ~~~ \\
	 R'(r,t) = {(t+\beta_0)}^{2/3};  &   \dot{R'}(r,t) =
	 { \mbox{$ 2 $}\over \mbox{$ 3 $} } {(t+\beta_0)}^{-1/3}.
	 \end{array}
 \right.
 \label{p15}
\end{equation}
The integration of the null geodesic (\ref{p5}) from $t = 0, \  r= 0$
till $t(r)$ results in
\begin{equation}
 3{(t+\beta_0)}^{1/3} = 3 {\beta_0}^{1/3} - r.
 \label{p16}
\end{equation}
With the solution (\ref{p16}) we can then write equations (\ref{p15})
along the null geodesic:
\begin{equation} 
 \left\{ \begin{array}{ll}
	 R={ \mbox{$ r $}\over \mbox{$ 9 $} } {(3 {\beta_0}^{1/3} - r)}^2;  &
	 \dot{R}=2r {( 3 {\beta_0}^{1/3} - r )}^{-1}; \\
	 ~~~  &  ~~~  \\
	 R'={ \mbox{$ 1 $}\over \mbox{$ 9 $} } {(3 {\beta_0}^{1/3} - r)}^2;  &
	 \dot{R'}=2 {( 3 {\beta_0}^{1/3} - r )}^{-1}.
	 \end{array}
 \right.
 \label{p17}
\end{equation}
Using the appropriate equation (\ref{p17}) and integrating equation
(\ref{p7}) from $I=0, \ r=0$ to $I(r)$, it is straightforward to show that
\begin{equation}
 1-I = {\left( \frac{3 {\beta_0}^{1/3} - r}{3 {\beta_0}^{1/3}} \right) }^2,
 \label{p18}
\end{equation}
and the redshift (\ref{p6}) parametrized along the geodesic becomes
\begin{equation}
 1+z= {\left( \frac{3 {\beta_0}^{1/3}}
			 {3 {\beta_0}^{1/3} - r}\right) }^2.
  \label{p19}
\end{equation}
If we consider equations (\ref{p14}) and (\ref{p19}), the luminosity
distance, number counts, volume and volume density are easily found 
to be
\begin{equation}
 d_l =  \frac{9r {\beta_0}^{4/3}}{{(3 {\beta_0}^{1/3}-r)}^2},
 \label{p20}
\end{equation}
\begin{equation}
 N_c =  \frac{2r^3}{9 M_G},
 \label{p21}
\end{equation}
\begin{equation}
 V = \frac{12 \pi r^3 {(3 \beta_0)}^4}{{(3 {\beta_0}^{1/3}-r)}^6},
 \label{p22}
\end{equation}
\begin{equation}
 \rho_v =  \frac{{(3 {\beta_0}^{1/3}-r)}^6}{54 \pi {(3 \beta_0)}^4} .
 \label{p23}
\end{equation}  

~~~~~As one can see the volume density here is expressed in terms of a
parameter along the geodesic.
Bonnor (1972) carried out a study bearing some similarities to the one
presented  in this paper, but with the major difference that here all
observational relations are calculated along the backward null cone, since
this is where astronomical observations are actually made. Bonnor's (1972)
volume density only applies to the present time hypersurface. Equation
(\ref{p23}) may change along the geodesic because it is a cumulative
density, averaging at bigger and bigger volumes with different local
densities.

~~~~~Finally, the local density (\ref{p4}) is given by
\begin{equation}
 \rho =  \frac{1}{6 \pi {(t+ \beta_0)}^2},
 \label{II1}
\end{equation}
and along the geodesic this equation becomes
\begin{equation}
 \rho = \frac{1}{6 \pi} {\left( \frac{3}{3 {\beta_0}^{1/3}-r}
        \right)}^6.
 \label{p24}
\end{equation}
It is important to stress that the presence of the coordinate $r$
in equation (\ref{p24}) does not mean a spatially inhomogenous local
density. The radial coordinate is only a parameter along the
geodesic and, due to the past null geodesic equation (\ref{p16}), each
value of $r$ corresponds to a single value of $t$. In other words, each $r$
corresponds to a specific $t =$ constant hypersurface given by equation
(\ref{p16}) and, hence, the density in equation (\ref{p24}) is
homogeneous, that is, constant at each such hypersurface. However,
equation (\ref{p24}) effectively changes along the geodesic as it goes
through different surfaces of $t$ constant and in this sense the model can be
thought of as inhomogeneous.

~~~~~An interesting consequence follows immediately from the results above.
If we use equation (\ref{p19}) and express $\rho_v$ in terms of the
redshift, we get
\begin{equation}
 \rho_v = \frac{1}{6 \pi {\beta_0}^2 {(1+z)}^3}
 \label{p25}
\end{equation}
which expanded in power series turns out to be
\begin{equation}
 \rho_v = \frac{1}{6 \pi {\beta_0}^2} \left( 1-3z+6z^2-10z^3+\ldots
 \right).
 \label{p26}
\end{equation} 
We can immediately see from equation (\ref{p26}) that for small
redshifts the volume density is constant, but as soon as $z$ increases
$\rho_v$ begins to depart from a constant value. Therefore, the
spatially homogeneous Einstein--de Sitter model does not appear to have
constant volume density as we go down the geodesic. We shall return to
this point later.

~~~~~As the function $\beta(r)$ determines the local time at which
$R~=~0$, the surface $t~+~\beta~=~0$ is a surface of singularity and,
hence, the physical region considered is given by the condition 
$t~+~\beta~>~0$. Considering the null geodesic (\ref{p16}) we can see that the
surface of singularity is at $r=3 {\beta_0}^{1/3}$ and the physical
region is $0 \leq r < 3 {\beta_0}^{1/3}$. Actually when $r \rightarrow 3
{\beta_0}^{1/3}$ the observational relations break down as
$z~\rightarrow~\infty, \ V~\rightarrow~\infty, \ d_l~\rightarrow~\infty,
\ \rho~\rightarrow~\infty, \ \rho_v~\rightarrow~0$.

~~~~~As one can see the volume density vanishes at the big bang, a
result that comes as a consequence of the definition adopted for 
$\rho_v$ in equation (\ref{p11}). At the big bang singularity hypersurface
the volume is infinite, but the total mass is finite. It is interesting to
note that one of the postulates of the so--called ``Pure Hierarchical
Models'' (Wertz 1970, p. 18) requires a vanishing global density as
the distance goes infinite, and a similar result was also conjectured by
Pietronero (1987) for a fractal distribution. This vanishing volume density
in a spatially homogeneous model might be interpreted as meaning that
the Einstein--de Sitter model is hierarchical at the asymptotic limit.

~~~~~In order to investigate a possible fractal behaviour in this model,
if we remember that $r \geq 0, \ \ d_l \geq 0$, equation (\ref{p20}) can
be inverted to get
\begin{equation}
 3{\beta_0}^{1/3}-r=3{\beta_0}^{1/3} { \left[ \frac{1}{2} + \sqrt{
                    \frac{d_l}{3 \beta_0} + \frac{1}{4} } \ \right]
		    }^{-1}.
  \label{pa1}
\end{equation}
With the equation above we can express the number count (\ref{p21}) and
the volume density (\ref{p23}) in terms of the luminosity distance.
These expressions may be written as
\begin{equation}
 N_c = \frac{6 \beta_0}{9 M_G} { \left[ 1- { \left( \frac{1}{2} + \sqrt{
       \frac{d_l}{3 \beta_0} + \frac{1}{4} }  \ \right) }^{-1}  \right]
       }^3,
  \label{pa2}
\end{equation} 
~~~~~
\begin{equation}
  \rho_v = \frac{1}{6 \pi {\beta_0}^2} { \left( \frac{1}{2} + \sqrt{
           \frac{d_l}{3 \beta_0} + \frac{1}{4} }  \ \right) }^{-6}.
    \label{pa3}
\end{equation} 
The power series expansions of equations (\ref{pa2}) and (\ref{pa3}) are
\begin{equation}
 N_c = \frac{2 {d_l}^3}{81 M_G {\beta_0}^2} 
       \left[ 1 - \frac{2}{\beta_0} d_l + \frac{3}{{\beta_0}^2} {d_l}^2
       - \frac{110}{27 {\beta_0}^3} {d_l}^3 + \ldots \right],
  \label{pa4}
\end{equation}
\begin{equation}
  \rho_v = \frac{1}{6 \pi {\beta_0}^2} \left[1-
	     \frac{2}{\beta_0} d_l + \frac{3}{{\beta_0}^2} {d_l}^2-
	     \frac{110}{27{\beta_0}^3} {d_l}^3+ \ldots \right].
  \label{p30}
\end{equation}
We can see again that $\rho_v$ is constant at the origin and remains
virtually unchanged for small values of $d_l$, a behaviour hardly in
line with the de Vaucouleurs' density power law (\ref{p13}). By
comparing equation (\ref{pa4}) with equation (\ref{p12}) we can also see
that at small values of the luminosity distance, the model has
fractal dimension equal to 3, that is, corresponding to a homogeneous
distribution, and this value changes as $d_l$ increases. Both results
are not in line with what one would expect if Einstein--de Sitter
had a single fractal behaviour.

~~~~~The results above can be seen graphically by plotting the logarithm
of $\rho_v$ against $d_l$. This method offers us an alternative
way of investigating whether Einstein--de Sitter has any fractal
behaviour. One cannot carry out this graphic investigation by means
of equation (\ref{p12}) because not only it does already assume a single
fractal behaviour for the dust, which is what one
wants to test, but also it involves the unknown constant $\sigma$. What
one can say is that if the distribution remains homogeneous throughout
the null geodesic, the volume density will not change and, therefore,
$\log \rho_v$ = constant and $D=3$. In this homogeneous situation the
plot of $\log \rho_v$ against $\log d_l$ would consist of a straight line
with zero slope.

~~~~~Figure 1 shows the log-log graph of $\rho_v$ plotted against
$d_l$ using their parametric relations (\ref{p23}) and (\ref{p20}) for
$0.001 \leq r \leq 1.5$. One can clearly see that for the luminosity
distance range $-1.2 \le \log d_l \le -1 \ (0.06 \stackrel{<}{\sim}
d_l \stackrel{<}{\sim} 0.1)$
the distribution starts to deviate significantly from a homogeneous
distribution. As at small values of $d_l$ the distribution is still
homogeneous, we can interpret the deviation as due to a significant
distancing from the initial time hypersurface. For $\log d_l \ge -0.8 \
(d_l \stackrel{>}{\sim} 0.16)$ on the observational relations are
 being evaluated at
increasingly very different and earlier epochs. The graph clearly  shows
that the Einstein--de Sitter model does not produce the results expected
from a fractal model. 

~~~~~It is also interesting to note that $\log d_l = -0.8 \ (r \cong
0.08)$ corresponds to $z \approx 0.04$ and that the deepest inhomogeneous
structures identified in the IRAS survey are at $ z \approx 0.07$
(Saunders et al. 1991). Therefore,
for the range where Einstein--de Sitter predicts homogeneity, the IRAS
survey does not find it and for the regions beyond the model starts to
deviate from it. If we accept these results at their
face value they might bring difficulties for a Einstein--de Sitter
universe as inhomogeneities are certainly present at least within 
$ z \approx 0.07$.

~~~~~~Recently, Efstathiou et al. (1991) have studied a deep survey
of faint blue galaxies with magnitudes around $24 - 26$. It would be of
interest to see how their sample and their modelling relates
to the model of this work. For this purpose, let us first find the redshift
distribution of the sources in this model. Considering equation
(\ref{p19}) we can write the number counts (\ref{p21}) along the geodesic
as a function of the redshift of the sources:
\begin{equation}
 N_c = \frac{6 \beta_0}{M_G} { \left[ 1- \frac{1}{\sqrt{1+z}} \right]
         }^3.
 \label{A30}
\end{equation}
The number of objects per unit redshift interval may be written as 
\begin{equation}
 \frac{d N_c}{d z} = \frac{9 \beta_0}{M_G} \frac{1}{{(1+z)}^{3/2}} {
                     \left[ 1- \frac{1}{\sqrt{1+z}} \right] }^2.
\label{A31}
\end{equation}
It is easy to see that equation (\ref{A31}) starts increasing from the
origin and then, after peaking at $z \cong 1.78$, begins to decrease
steadily, without a sharp fall.

~~~~~Although it is not obvious how equation (\ref{A31}) relates to the
data of Efstathiou et al. (1991), especially because of the uncertainty of
the redshift distribution for their faint galaxies, we can at least say
that one of the redshift distributions which they considered ``realistic''
has some functional similarities to equation (\ref{A31}), in the sense that it
also starts increasing and then  decreases steadily, without sharp cutoffs,
after a maximum. Nonetheless, we have to bear in mind that the calculations
shown here assume that the sources have uniform intrinsic luminosity, and
that means that specific redshifts correspond to specific luminosity
distances and apparent magnitudes. This is certainly a simplification as
in real astronomical observations one has objects of different magnitudes
at equivalent redshifts, and vice-versa. To be able to infer the
objects' distance from apparent magnitudes, one needs a model based on
their intrinsic physical processes that gives intrinsic luminosity. In
other words, one must have a model for galactic evolution or, perhaps
more specifically, galactic color evolution. We shall not consider here the
problem of linking equation (\ref{A31}) to real astronomical data
acquisition methods and galactic evolution models.

~~~~~Before the end of this section, let us briefly discuss the claim of
Efstathiou et al. (1991), based on calculations of the two-point angular
correlation function, that faint blue galaxies are weakly clustered. The
important point is that if the galactic distribution
forms a fractal structure, it would not necessarily be easily detected by the
angular correlation function. This point was highlighted by Coleman \&
Pietronero (1991) who measured angular correlations of simulations of
three dimensional fractal structures and found out that they have a
tendency to homogenize at angles which approach a fraction of the sample
angle. This is explained by the peculiar properties of fractals whose
dimensions may change once they are projected. Thus, it would appear that
conclusions about homogeneity based on the current angular
correlation analysis are open to controversy. Fractals are basically
simple, but they are also subtle and therefore can be elusive. 

\section{Conclusion}
  ~~~~~In this work the theory developed in Ribeiro (1992) for a
  relativistic hierarchical (fractal) cosmology has been applied to the
  special case of the Einstein--de Sitter model. We have calculated the
  observational relations for this particular model and studied whether it
  is compatible with a fractal description of galactic clustering. The
  results show that the model under consideration does not
  possess fractal features along the backward null cone and
  fails to predict the inhomogeneities observed at the scale of the IRAS
  survey.  Besides, although the model is homogeneous, it does
  not appear to remain so along the past light cone. We have
  also found a vanishing volume density at the big bang. Finally,
  we have obtained the equation for redshift distribution of the sources in
  a Einstein-de Sitter model and discussed the possible relationship of
  this theoretical distribution with the clustering analysis of a recent
  survey of deep faint galaxies. 

  ~~~~~~The inhomogeneity of the Einstein-de Sitter model is a
  consequence of the manner densities are expressed here. The volume
  density $\rho_v$ is measured along the past null geodesic which goes
  through hypersurfaces of $t$ constant, where each one has different
  values for the proper density. It changes along the geodesic
  because it is a cumulative density, averaging at bigger and bigger
  volumes, in a way that adds more and more different local densities
  of each spatial section of the model. The proper density $\rho$ also
  changes when expressed along the geodesic and for the same physical
  reason. Hence, in this sense the ``homogeneity'' of the Einstein-de
  Sitter model does not survive.

  ~~~~~The analysis shown above is an example of the problems of
  applying measures of ``homogeneity'' even to a universe model that
  really is spatially homogeneous. It is clear that a ``homogeneous''
  model may be taken to be inhomogeneous, depending on how we look at it.
  Furthermore, as the model apparently has difficulties in accounting for
  the observed inhomogeneities, that may compel us to conjecture that
  the Einstein--de Sitter model could end up being a victim of its own
  strength: it is too simple. From a relativistic fractal cosmological
  point of view what is really needed are solutions of Einstein's field
  equations with fractal behaviour along the past null cone. We intend
  to deal with such solutions in paper III.

  \vspace{30mm}
  \begin{flushleft}
  {\large \bf Acknowledgements}
  \end{flushleft}
  \vspace{5mm}

  ~~~~~It is my pleasure to thank W. B. Bonnor for discussions and many
  suggestions which led to this paper and enriched it. I am also grateful
  to M. A. H.  MacCallum for numerous valuable discussions and
  suggestions, S. Oliver and P. Coles for discussions and the referee for
  helpful comments. This work had the financial support of Brazil's
  Ministry of Education agency CAPES.

\vspace{15mm}       
\begin{flushleft}
{\large \bf References}
\end{flushleft}
  \begin{description}
    \item  Bonnor, W. B. 1972, M. N. R. A. S., 159, 261.
    \item  Coleman, P. H., \& Pietronero, L. 1991, preprint.          
    \item  Efstathiou, G. et al. 1991, Ap. J. (Letters), 380, L47.    
    \item  Jones, B. J. T., et al. 1988, Ap. J. (Letters), 332, L1.
    \item  de Lapparent, V., Geller, M. J., \& Huchra, J. P. 1986,
           Ap. J. (Letters), 302, L1.
    \item  Mart\'{\i}nez, V. J., et al. 1990, Ap. J., 357, 50.
    \item  Pietronero, L. 1987, Physica, 144A, 257.
    \item  Ribeiro, M. B. 1992, Ap. J., 388, 1 (paper I).
    \item  Saunders, W., et al. 1991, Nature, 349, 32.
    \item  Tolman, R. C. 1934, Proc. Nat. Acad. Sci. (Wash.), 20, 169.
    \item  Wertz, J. R. 1970, ``Newtonian Hierarchical Cosmology'',
           PhD thesis (University of Texas at Austin).
  \end{description}
\newpage
\begin{center}
 {\large Figure Caption}
 \vspace{10mm}
\end{center}
\begin{description}
 \item 
 Log-log plot of volume density $\rho_v$ given by
   equation (\ref{p23}) against the luminosity distance $d_l$ given by
   equation (\ref{p20}) in the range $0.001 \leq r \leq 1.5$ and with
   $\beta_0=2.7$ (units are geometrical with $c=G=1$, distance is given
   in Gpc and the Hubble constant assumed to be 75 km/s/Mpc). We can
   clearly see that in the Einstein--de Sitter model the distribution
   does not appear to remain homogeneous along the geodesic and the fractal
   dimension departs from the initial value~3.
\end{description}
\begin{figure}
\includegraphics[width=12cm,angle=-90]{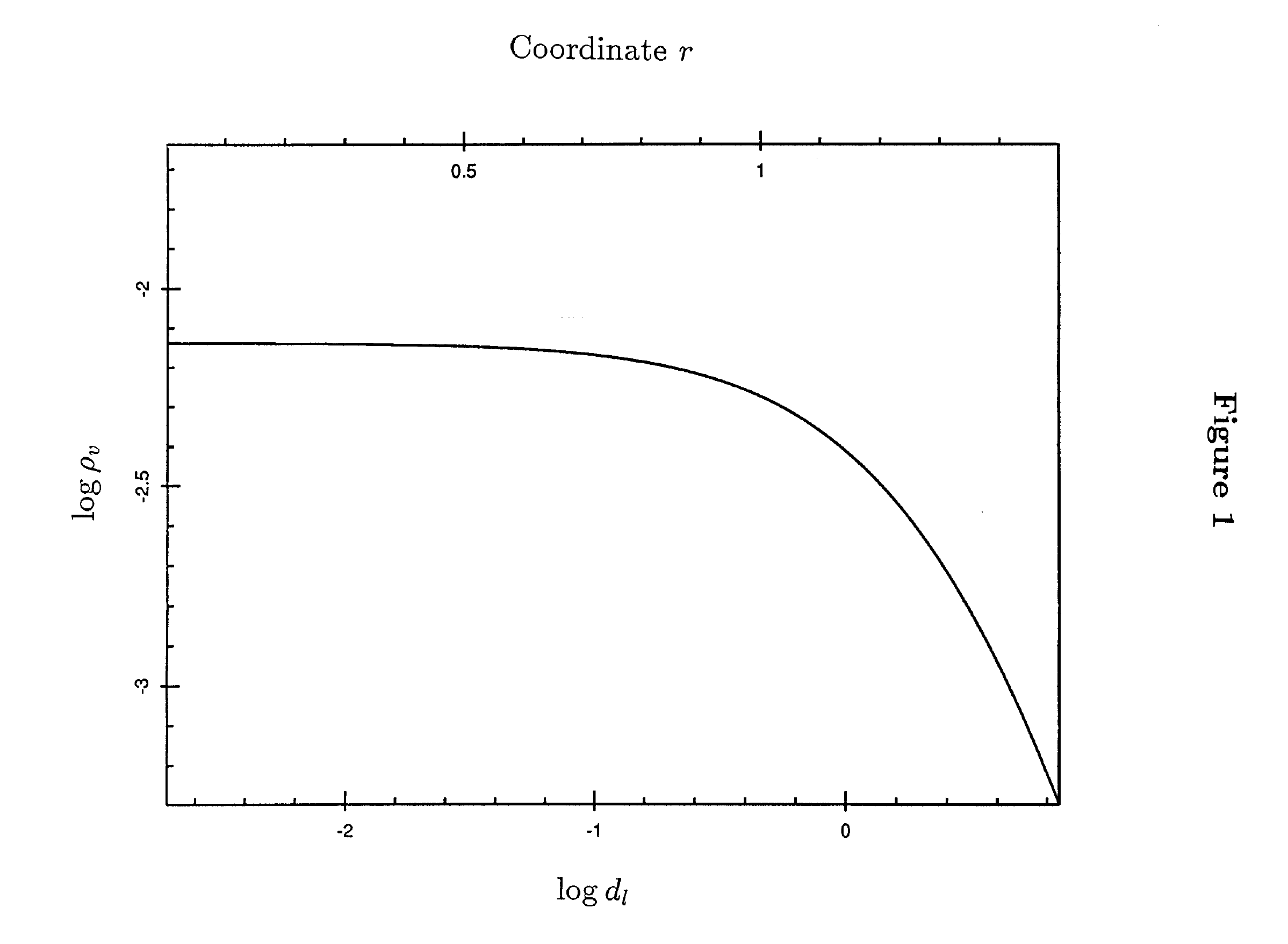}
\end{figure}
\end{document}